# A new paradigm for parallelized STED microscopy

Omer Wagner,[1] Ori Cheshnovsky,[2]* Yael Roichman[2],*

[1]*School of Engineering, Tel-Aviv University, Tel-Aviv, Israel*
[1]*School of Chemistry, Tel-Aviv University, Tel-Aviv, Israel*
*Corresponding authors: orich@post.tau.ac.il, roichman@tau.ac.il

We propose a new design for a multi-color, parallelized STED microscope, capable of multiple beam scanning. Our design is based on a common programmable diffracting optical element used to split, shape, and align both the excitation beams as well as the depletion beams, and a CCD camera to image florescence from all scanning beams simultaneously. Here we describe a prototype implementation of a two beam STED based on our novel design, which could easily be generalized to multi-color STED, with up to 100 parallel scanning beams given enough laser power, and high rate STED imaging with the use of a high-speed response multianode photomultiplier detector.

Super-resolution far field microscopy has reached beyond Abbe's diffraction limit down to resolutions of approximately 10nm using diverse methods [1–4]. Stimulated emission depletion microscopy (STED) [3,5,6], stands out due to its ability to create high resolution, multiple-color image sequences at video rates, provided that the scan area is small and suitable fluorophores are chosen. These features make STED a desirable imaging technique for live cells [7].

In STED the resolution improvement is achieved by a deterministic pattern of excitation and de-excitation of fluorophores. A diffraction limited excitation source (Gaussian shaped) is projected together with a co-centered depletion doughnut shaped laser beam. The doughnut beam forces molecules located at the periphery of the focal center to become temporarily non-fluorescent by inducing stimulated emission in the depletion wavelength. A key element in STED is the non-linear dependence of the de-excitation on the laser intensity, which enables its saturation. Hence, one can reduce the volume of emitting molecules by increasing the power of the depletion doughnut beam. By nanometric precision scanning of this sub-diffraction limit volume, super-resolution imaging is achieved. With this approach, lateral and axial resolution of 10nm and 40nm, respectively, were demonstrated.

Video rate STED imaging is available for using high response rate photomultipliers for a field of view of approximately $2\times2\mu m^2$ [1,8]. One way to improve STED imaging rate is by using parallel scanning, which requires generation of well aligned arrays of excitation and depletion beams. Existing parallel-beam STED implementations [9], add complexity to the STED setup, are not easily scalable, and are restricted spectrally. In this letter we propose a new design principle for STED microscopy, greatly simplifying the optical setup and significantly increasing the field of view scan velocity. Our approach is based on a programmable spatial light modulator (SLM) which is used as a common beam shaping element, for the STED's excitation and depletion laser sources. The same optical element is also used to split the beams to enable parallelized scanning. The programmable nature of the SLM allows us to generate an array of STED beams with any two colors of choice, and to correct for optical train aberrations in real time.

Our implementation consists of an excitation laser (450nm CW, z-laser Z20M18H-F-450-pe 20 mW), and a STED laser (532nm CW, Verdi v6 coherent 6W) coupled to the same optical path using a dichroic mirror. Both beams are expanded to overfill a spatial light modulator (SLM, Hamamatsu X-10468-04) where they are phase modulated with a computer-generated phase hologram.

The resulting hologram is scanned with a piezo controlled rotating mirror (PI, S-330), contracted and imaged on the back aperture of a microscope objective (100x 1.4 NA). A quarter-wave plate in the back aperture of the microscope objective ensured circular polarization of both beams. Sample fluorescence was filtered, and recorded on a CCD camera (Photometrics CoolSNAP HQ2).

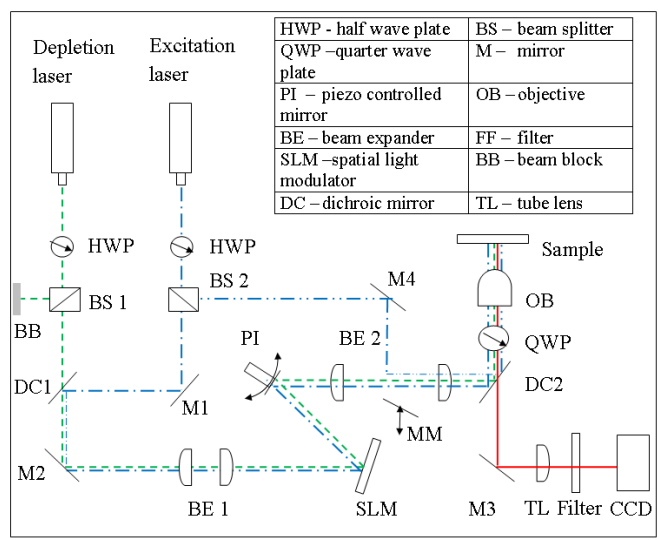

Fig. 1. (Color online) SLM based parallelized STED. An excitation laser beam (450nm) and a STED beam (532nm), with vertical polarization, are coupled (DC1), expanded (BE1), and projected onto a Hamamatsu X-10468-04 SLM. The beams are reflected from a piezo controlled rotating mirror (PI), and coupled into a microscope using relay optics (BE2) and a dichroic mirror (DC2).

We use multi-wavelength holography [10] to generate the excitation and depletion light pattern with a single hologram programmed onto an SLM. Our approach is to design a hologram which is comprised of two planes, one encoding the excitation pattern and the other the depletion pattern. Both laser beams produce the same, undistorted, 3D image after interaction with the SLM [10,11] differing in position along the optical axis and in scale. By choosing the location of the projected hologram and the distance between the two image planes we can ensure that the desired excitation and depletion beam patterns will align on a single plane (Fig 2a,b). In order to observe the fluorescence via the infinity corrected optics of the microscope, it is essential that this plane will coincide with the focus plane of the microscope. This is achieved by rendering the beams slightly converging by tuning the beam expander (BE1, Fig. 1), which also enables us to shift the undiffracted light away from the focal plane [12]. Using this method, we are able to simultaneously split the excitation beam and the depletion beam into a grid of $N$ STED couples (Fig 2c, $N=8$) increasing our field of view $N$-fold. In principle, the number of parallel scanning beams is limited by our field of view to about 100 beams, and by the total laser power the SLM can withhold. For STED applications this total laser power imposes a restriction on the effective number of parallel STED points that can be scanned, since the resolution is depended on the power of each depletion beam and the saturation intensity, $I_s$, of the fluorophores. We measured experimentally that about a 100mW of laser power at the sample is needed for each STED depletion beam to acquire 150nm resolution on patterns were characterized, calibrated, and aligned using volumetric imaging [14], combined with adaptive optics [12,15] (Fig. 2b and c), in contrast with former calibration protocols that used scattering from nano gold particles for that purpose [16]. Scanning calibration is then done by applying particle tracking algorithms [17] to locate the two beams' center and their displacement as a function of stirring voltage. At this stage basic adaptive optics algorithms were used to counteract geometrical aberrations in the optical train [15].

In our implementation we used 60nm microspheres (Bangs Laboratories Inc., P(S/R-NH2), Lot. 9082) dyed with atto425 fluorescent dye (ATTO-TEC GmbH). Samples were prepare following the procedure of [18]. Data from several consecutive scans was compared; each scan consisting of an initial scan of a single excitation beam and a second scan of two beams including both excitation and depletion beam. Figure 3 shows a typical result of the double scan; Fig 3a, an image of a particle as seen with a single excitation beam, and Fig 3b, showing the same particle imaged by STED. A projection of a cross-section along the bead (Fig. 3c) shows an improved resolution from 221nm (confocal imaging) to 175nm (STED imaging), at a depletion beam power of 88 MW/cm$^2$ on the sample. Applying additional filtering, such as *image scanning microscopy* algorithms [19], we can further improve our resolution to a value of 153nm (Fig.3c,d).

We have tested our parallel STED imaging concept with a dual beam configuration. Here both beams are scanned over a field of view of 1.29x6.11 microns with an polystyrene colloids dyed with Atto425. This power also is consistent with previous reports [13]. The resulting light overlap of 1.03x1.22 microns at the center portion of the image, where a particle is located (Fig. 4). Clearly, the PSFs of the two beams are different, as seen from the difference in the particle shape which was obtained by the two beams (compare Fig. 4a and b). Stitching both images to form a larger field of view is done based on the calibrated known position of the two STED beams, as demonstrated in Fig. 4c and d. The good coincidence of the two images indicates the validity and accuracy of our approach.

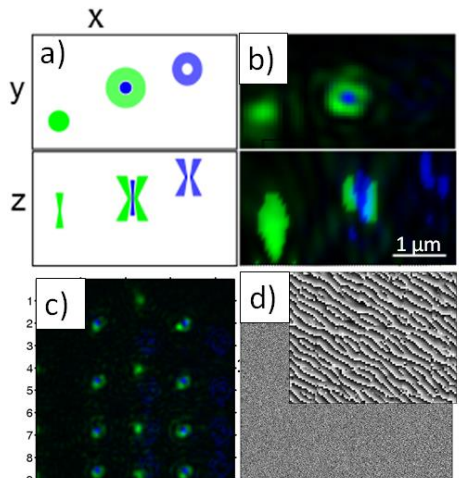

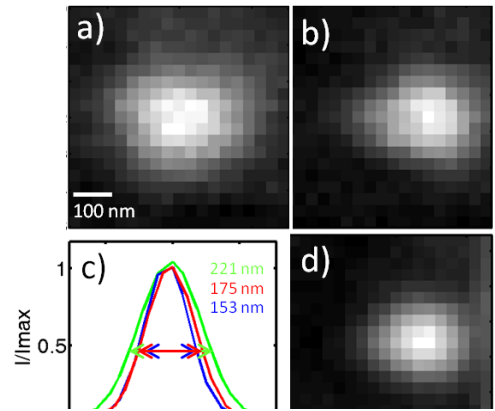

Fig. 2. (Color online) Light pattern generation using multi-wavelength holography. A hologram projecting two patters at different locations is projected. a) Design; XY and XZ projections of the hologram reconstructed by two lasers, made to coincide in a single plane. b) XY and XZ projections of the light pattern of excitation and depletion beams (falsely colored), measured by volumetric imaging [14]. c) XY projection of a measured 8 beam STED light pattern (falsely colored), created using the phase hologram in (d). Inset shows a part of the phase hologram magnified.

Fig. 3. (Color online) Comparison between confocal and STED imaging of single 60nm dyed polystyrene particle. a) Image of a single particle obtained by scanning a single excitation beam (450nm). b) Image of the same bead obtained by scanning in STED configuration (450nm at 5 kW/cm$^2$, 532nm at 88 MW/cm$^2$). d) STED image of the bead refined with structured illumination algorithm [19] c) Cross sections of image (a) (green), (b) (red) and (d) (blue).

In this paper we introduced a novel method to achieve parallel STED microscopy with the potential of surpassing video rate in super-resolution imaging. We have implemented our idea in a proof-of-principle setup with two parallel scanning beams and achieved resolutions down to 155nm ± 10nm at 0.5 Hz. These values can be greatly improved with some hardware modification, the most significant of which is the use of a high-speed, sensitive 8x8 multianode photomultiplier instead of a 10Hz air cooled CCD camera. The high sensitivity of the multianode photomultiplier will significantly enhance signal to noise ratio during imaging and its fast response will reduce bleaching effects, which translates to increased spatial resolution and a reduction of laser power needed for depletion. In this configuration scanning in parallel 32 beams with 1μs dwell time is feasible. Pinholing will be achieved by adjusting the field of view to 20x20μm, each detector in the anode array receiving illumination from a 2.5x2.5 μm window. Arranging the parallel beams in a 4x8 array will ensure rejection of illumination originating from out-of-plane beams. Scanning in steps of 30nm an area of 2.5x5 μm will result in an imaging rate of 72 fps of a 20x20 μm field of view, which is well beyond current capabilities.

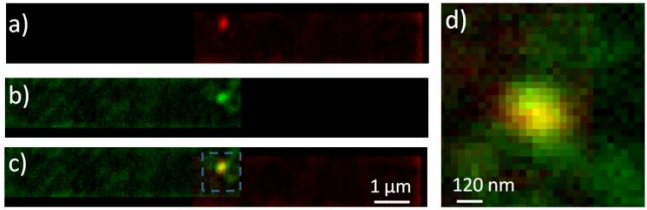

Fig. 4. (Color online) Two beam parallel STED. a) An image formed by the left beam. b) An image formed by the right beam. The scan area of both beams overlapped in the center of the field of view where a single particle is located. c) Stitching together the two images to increase the field of view approximately two fold. d) The stitching area magnified, red represents the left beam, green the right beam and yellow, the area of overlap, demonstrating the accuracy of image stitching.

Using the same hologram to create both the excitation pattern as well as the depletion pattern means that only half of the lasers' power is used for imaging, and close out-of-plane beams exist near each scanned beam. These constraints not only reduce the parallelization level achievable with a given laser, but also affect the minimal distance possible between adjacent detector anodes. Since the power of the depletion laser is the restricting parallelization factor, it is possible to increase parallelization level by increasing the fraction of beam power directed to the depletion pattern by the SLM [20]. The efficiency of our approach can be further increased by using two SLMs, one in charge of the excitation beam and one in charge of the depleting beam. . In this way the restriction on distance between detector anodes will be lifted as well. These improvements will come at the expense of stable excitation and depletion beam alignment as exists in the regular STED scheme.

The new design proposed here can be readily adapted to three dimensional scans, changing the doughnut mode of the depletion beam to the standard 3D shape used in conventional STED [16]. The use of multi-wavelength holography for STED can enhance various aspects of STED system design, such as self-alignment [21], and simplify multi-color STED imaging [22].

We would like to acknowledge Tal Maya for her helpful contribution to this work, and Dr. Yuval Ebenstein for insightful conversations. This research was supported by the James Franck German-Israeli Binational Program in Laser-Matter interactions.

1. D. Toomre and J. Bewersdorf, Annu. Rev. Cell Dev. Biol. **26**, 285 (2010).
2. A. Diaspro, *Nanoscopy and Multidimensional Optical Fluorescence Microscopy* (CRC Press/Taylor & Francis Group, Boca Raton, 2010).
3. S. W. Hell, Science **316**, 1153 (2007).
4. N. G. Walter, C.-Y. Huang, A. J. Manzo, and M. A. Sobhy, Nat Meth **5**, 475 (2008).
5. T. Grotjohann et al., Nature (2011).
6. S. W. Hell and J. Wichmann, Optics Letters **19**, 780 (1994).
7. S. Saka and S. O. Rizzoli, BioEssays : News and Reviews in Molecular, Cellular and Developmental Biology **34**, 386 (2012).
8. M. A. Lauterbach et al., Journal of Biophotonics **3**, 417 (2010).
9. P. Bingen, M. Reuss, J. Engelhardt, and S. W. Hell, Optics Express **19**, 23716 (2011).
10. S.-H. Lee and D. G. Grier, **13**, 7458 (2005).
11. N. Zhang et al., Applied Optics **49**, 2456 (2010).
12. M. Polin et al., Opt. Express **13**, 5831 (2005).
13. B. R. Rankin, R. R. Kellner, and S. W. Hell, Optics Letters **33**, 2491 (2008).
14. Y. Roichman, I. Cholis, and D. G. Grier, Opt. Express **14**, 10907 (2006).
15. Y. Roichman, A. Waldron, E. Gardel, and D. G. Grier, Applied Optics **45**, 3425 (2006).
16. K. I. Willig, B. Harke, R. Medda, and S. W. Hell, Nat. Methods **4**, 915 (2007).
17. J. C. Crocker and D. G. Grier, **310**, 298 (1996).
18. C. A. Wurm et al., **591**, 185 (2010).
19. C. B. Müller and J. Enderlein, Phys. Rev. Lett. **104**, 1 (2010).
20. J. E. Curtis, B. A. Koss, and D. G. Grier, **207**, 169 (2002).
21. D. Wildanger et al., Opt. Express **17**, 16100 (2009).
22. E. Auksorius et al., Optics Letters **33**, 113 (2008).
1. D. Toomre and J. Bewersdorf, "A New Wave of Cellular Imaging," Annu. Rev. Cell Dev. Biol. **26**, 285–314 (2010).

2. A. Diaspro, *Nanoscopy and Multidimensional Optical Fluorescence Microscopy* (CRC Press/Taylor & Francis Group, 2010).
3. S. W. Hell, "Far-Field Optical Nanoscopy," Science **316**, 1153–1158 (2007).
4. N. G. Walter, C.-Y. Huang, A. J. Manzo, and M. A. Sobhy, "Do-it-yourself guide: how to use the modern single-molecule toolkit," Nat Meth **5**, 475–489 (2008).
5. T. Grotjohann, I. Testa, M. Leutenegger, H. Bock, N. T. Urban, F. Lavoie-Cardinal, K. I. Willig, C. Eggeling, S. Jakobs, and S. W. Hell, "Diffraction-unlimited all-optical imaging and writing with a photochromic GFP," Nature (2011).
6. S. W. Hell and J. Wichmann, "Breaking the diffraction resolution limit by stimulated emission: stimulated-emission-depletion fluorescence microscopy.," Optics letters **19**, 780–2 (1994).
7. S. Saka and S. O. Rizzoli, "Super-resolution imaging prompts re-thinking of cell biology mechanisms: selected cases using stimulated emission depletion microscopy.," BioEssays : news and reviews in molecular, cellular and developmental biology **34**, 386–95 (2012).
8. M. A. Lauterbach, J. Keller, A. Schönle, D. Kamin, V. Westphal, S. O. Rizzoli, and S. W. Hell, "Comparing video-rate STED nanoscopy and confocal microscopy of living neurons," Journal of Biophotonics **3**, 417–424 (2010).
9. P. Bingen, M. Reuss, J. Engelhardt, and S. W. Hell, "Parallelized STED fluorescence nanoscopy.," Optics express **19**, 23716–26 (2011).
10. S.-H. Lee and D. G. Grier, "Robustness of holographic optical traps against phase scaling errors," **13**, 7458–7465 (2005).
11. N. Zhang, J. a. Davis, I. Moreno, J. Lin, K.-J. Moh, D. M. Cottrell, and X. Yuan, "Analysis of fractional vortex beams using a vortex grating spectrum analyzer," Applied Optics **49**, 2456 (2010).
12. M. Polin, K. Ladavac, S. H. Lee, Y. Roichman, and D. G. Grier, "Optimized holographic optical traps," Opt. Express **13**, 5831–5845 (2005).
13. B. R. Rankin, R. R. Kellner, and S. W. Hell, "Stimulated-emission-depletion microscopy with a multicolor stimulated-Raman-scattering light source.," Optics letters **33**, 2491–3 (2008).
14. Y. Roichman, I. Cholis, and D. G. Grier, "Volumetric imaging of holographic optical traps.," Opt. Express **14**, 10907–12 (2006).
15. Y. Roichman, A. Waldron, E. Gardel, and D. G. Grier, "Optical traps with geometric aberrations.," Applied optics **45**, 3425–9 (2006).
16. K. I. Willig, B. Harke, R. Medda, and S. W. Hell, "STED microscopy with continuous wave beams," Nat. Methods **4**, 915–918 (2007).
17. J. C. Crocker and D. G. Grier, "Methods of Digital Video Microscopy for Colloidal Studies," **310**, 298–310 (1996).
18. C. A. Wurm, D. Neumann, R. Schmidt, A. Egner, and S. Jakobs, "Sample Preparation for STE Microscopy," **591**, 185–199 (2010).
19. C. B. Müller and J. Enderlein, "Image Scanning Microscopy," Phys. Rev. Lett. **104**, 1–4 (2010).
20. J. E. Curtis, B. A. Koss, and D. G. Grier, "Dynamic holographic optical tweezers," **207**, 169–175 (2002).
21. D. Wildanger, J. Bückers, V. Westphal, S. W. Hell, and L. Kastrup, "A STED microscope aligned by design," Opt. Express **17**, 16100 (2009).
22. E. Auksorius, B. R. Boruah, C. Dunsby, P. M. P. Lanigan, G. Kennedy, M. a a Neil, and P. M. W. French, "Stimulated emission depletion microscopy with a supercontinuum source and fluorescence lifetime imaging.," Optics letters **33**, 113–5 (2008).